\documentclass{amsart}
\usepackage{amsmath,amsfonts,amssymb,amsthm}
\usepackage[english]{babel}
\usepackage{graphicx}
\usepackage{url}
\usepackage[round]{natbib}
\newcommand{\forarxiv}[1]{#1}
\newcommand{\notforarxiv}[1]{}

\newcommand{\vwork}{\delta}
\newcommand{\EE}{\mathbb{E}}
\newcommand{\calA}{\mathcal{A}}
\newcommand{\calB}{\mathcal{B}}
\newcommand{\calE}{\mathcal{E}}
\newcommand{\calV}{\mathcal{V}}
\newcommand{\bubbles}{\mathcal{B}}
\newcommand{\ADCL}{\operatorname{ADCL}}
\newcommand{\troot}{\mathsf{\rho}}

\newcommand{\RMP}{\operatorname{RMP}}
\newcommand{\RMD}{\operatorname{RMD}}
\newcommand{\maxleaves}{k}

\newcommand{\cddlib}{\textsf{cddlib}}

\newcommand{\proxmass}{\pi}
\newcommand{\wksubtot}{\omega}
\newcommand{\cldist}{\chi}
\newcommand{\bublen}{\lambda}
\newcommand{\bubmass}{\mu}
\newcommand{\wkdstl}{\alpha}
\newcommand{\wkprox}{\beta}
\newcommand{\leafset}{X}
\newcommand{\PS}{\varphi}
\newcommand{\PSset}{\psi}

\newcommand{\pplacer}{\textsf{pplacer}}

\newcommand{\rppr}{\textsf{rppr}}

\newcommand{\minadcl}{\textsf{min\_adcl}}
\newcommand{\minadcltree}{\textsf{min\_adcl\_tree}}

\newtheorem{lem}{Lemma}

\newtheorem{prop}{Proposition}

\newtheorem{prob}{Problem}
\newtheorem{defn}{Definition}

\newtheorem{alg}{Algorithm}

\newcommand{\eat}[1]{}

\newcommand{\FIGmassTransport}{\
\begin{figure}[ht]
\begin{center}
  \forarxiv{\includegraphics[width=13cm]{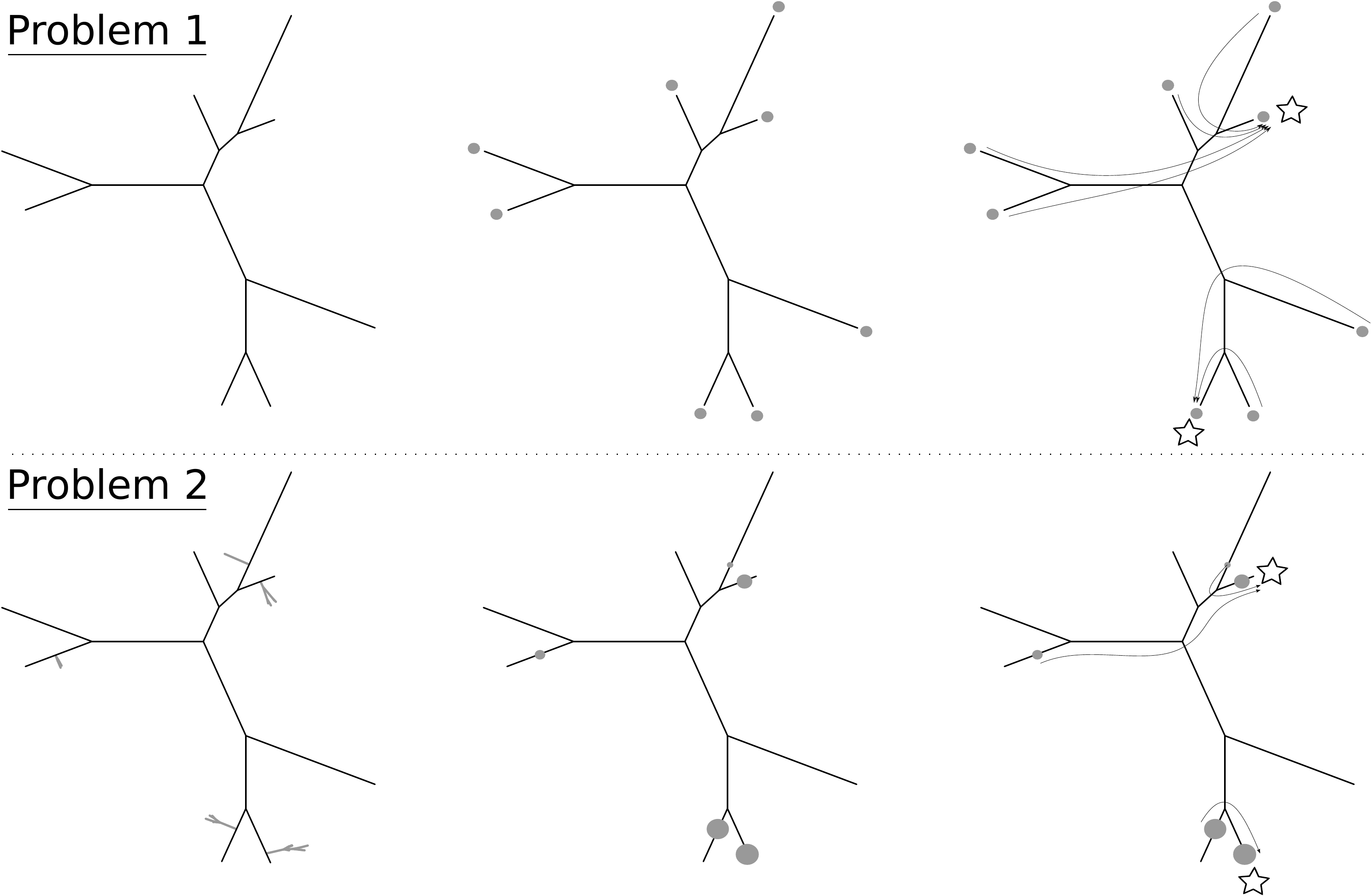}}
\end{center}
\caption{\
  A diagram showing two example Average Distance to the Closest Leaf (ADCL) minimization problems.
  The $k$ selected leaves are marked with hollow stars; in this case $k=2$.
  Problem 1 is to minimize the average distance from each leaf to its closest selected leaf.
  Problem 2 is to minimize the average distance from the query sequences (gray branches) to their closest reference sequence (reference sequence subtree in black).
  Both of these problems can be thought of as instances of Problem 3, which is to minimize the work required to move mass (gray circles) to a subset of $k$ leaves.
  In Problem 1, a unit of mass is uniformly divided amongst the leaves of the tree.
  In Problem 2, mass is distributed in proportion to the number of query leaves that attach at that point.
}
\label{FIGmassTransport}
\end{figure}
}
\newcommand{\refFIGmassTransport}{1}

\newcommand{\FIGlocalMinimum}{\
\begin{figure}[ht]
\begin{center}
  \forarxiv{\includegraphics[width=6cm]{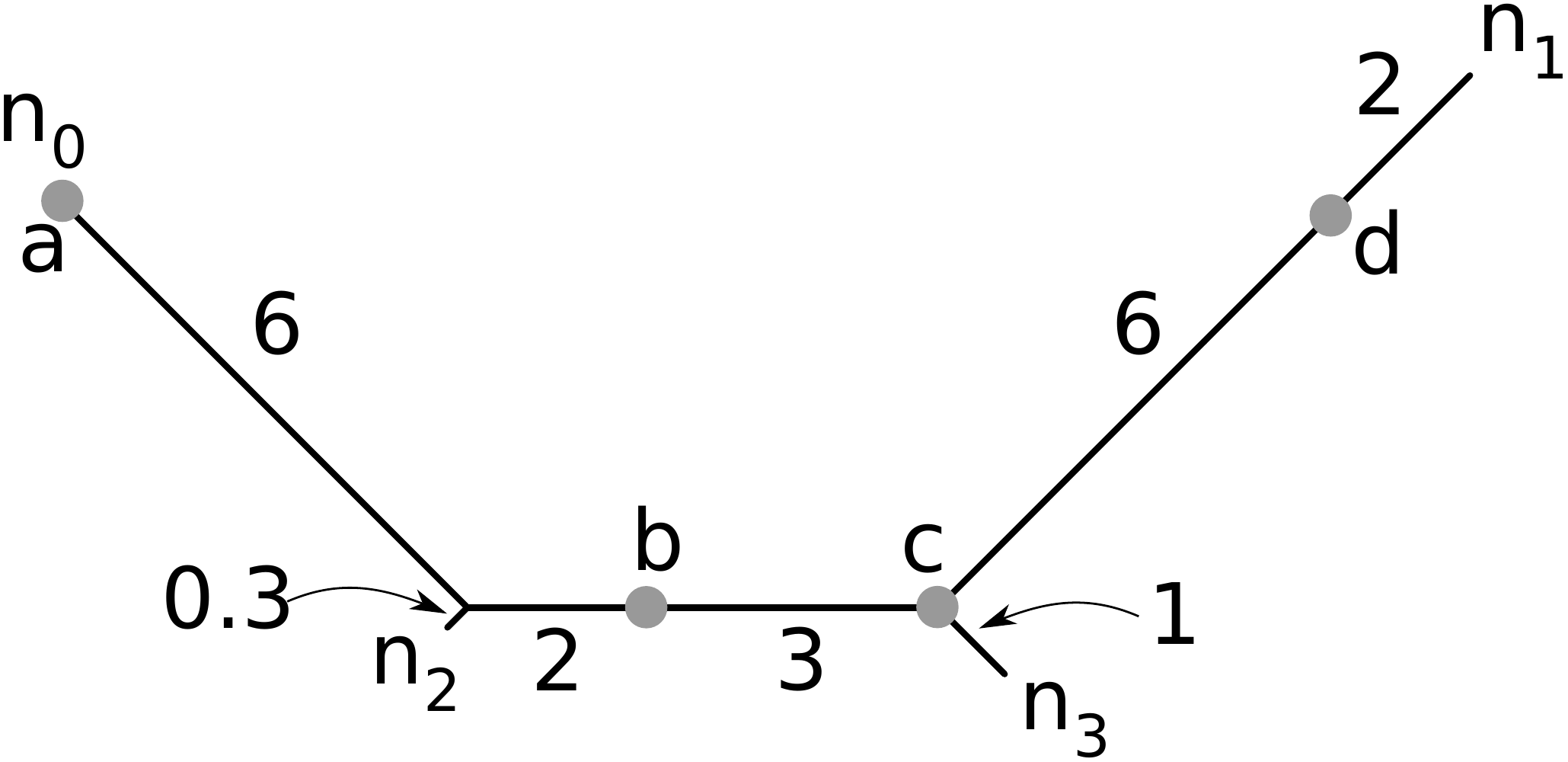}}
\end{center}
\caption{\
  An example where the Partitioning Among Medoids (PAM) algorithm gets stuck in a local minimum.
  Assume masses of equal magnitude $a,\dots,d$ on a tree with leaves $n_0,\dots,n_3$, and two leaves are desired from the algorithm (i.e. $k=2$).
  Branch lengths are as marked on the tree.
  The optimal solution is to take $\{n_0,n_3\}$.
  However, if PAM starts with $\{n_1,n_2\}$, it will not be able to make it to the optimal solution by changing one leaf at a time because the ADCL values for every other pair is greater than that for $\{n_0,n_3\}$ and $\{n_1,n_2\}$ (Table~\refTABlocalMinimum).
}
\label{FIGlocalMinimum}
\end{figure}
}
\newcommand{\refFIGlocalMinimum}{2}

\newcommand{\FIGrmdRmp}{\
\begin{figure}[ht]
\begin{center}
  \forarxiv{\includegraphics[width=6cm]{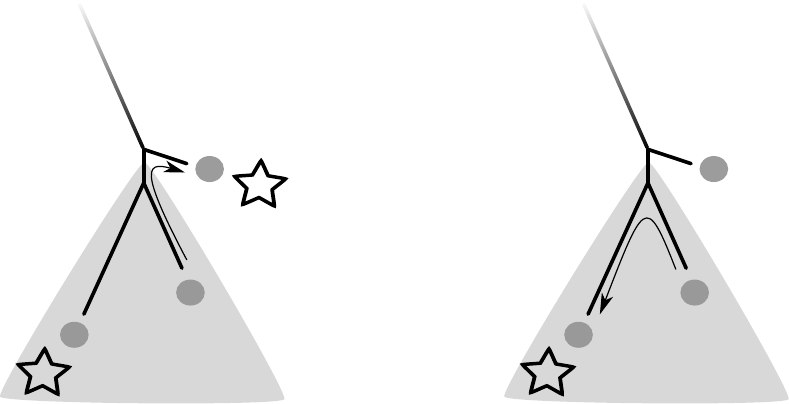}}
\end{center}
\caption{\
  The movement of mass within a subtree depends on the selection of leaves outside the subtree, motivating a dynamic program that keeps solutions that could be optimal for a variety of circumstances outside of the tree.
Here, stars represent selected leaves and filled circles represent masses.
}
\label{FIGrmdRmp}
\end{figure}
}
\newcommand{\refFIGrmdRmp}{3}

\newcommand{\FIGlowerHull}{\
\begin{figure}[ht]
\begin{center}
  \forarxiv{\includegraphics[width=6cm]{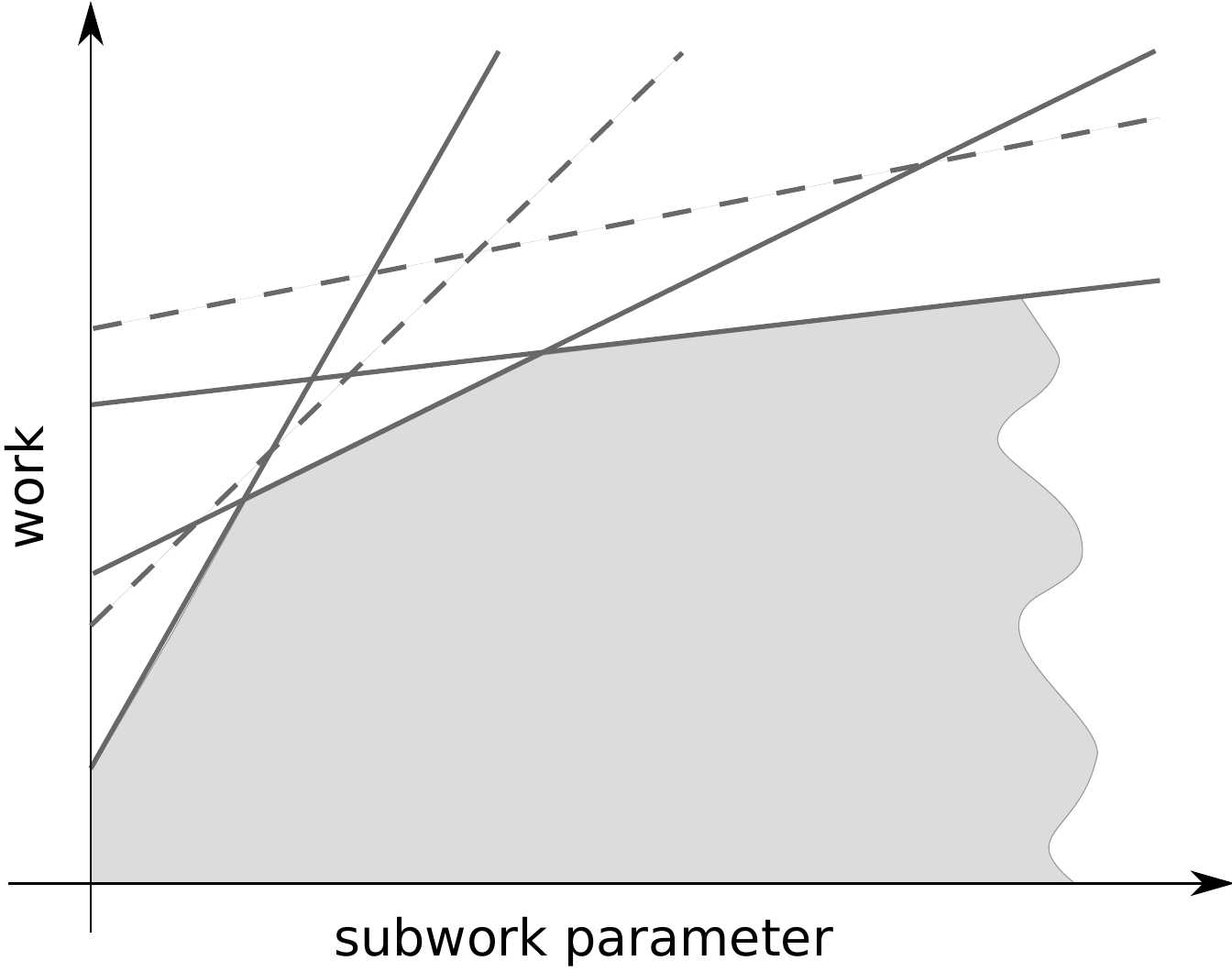}}
\end{center}
\caption{\
  A visual depiction of the method of removing partial solutions that could not be optimal for any setting in the rest of the tree.
  Each line represents the total work for a partial solution that has subwork parameter $x$.
  Because the dashed lines are not minimal for any value of $x$, they can be discarded.
}
\label{FIGlowerHull}
\end{figure}
}
\newcommand{\refFIGlowerHull}{4}

\newcommand{\FIGpamOptimality}{\
\begin{figure}[ht]
\begin{center}
  \forarxiv{\includegraphics[height=6cm]{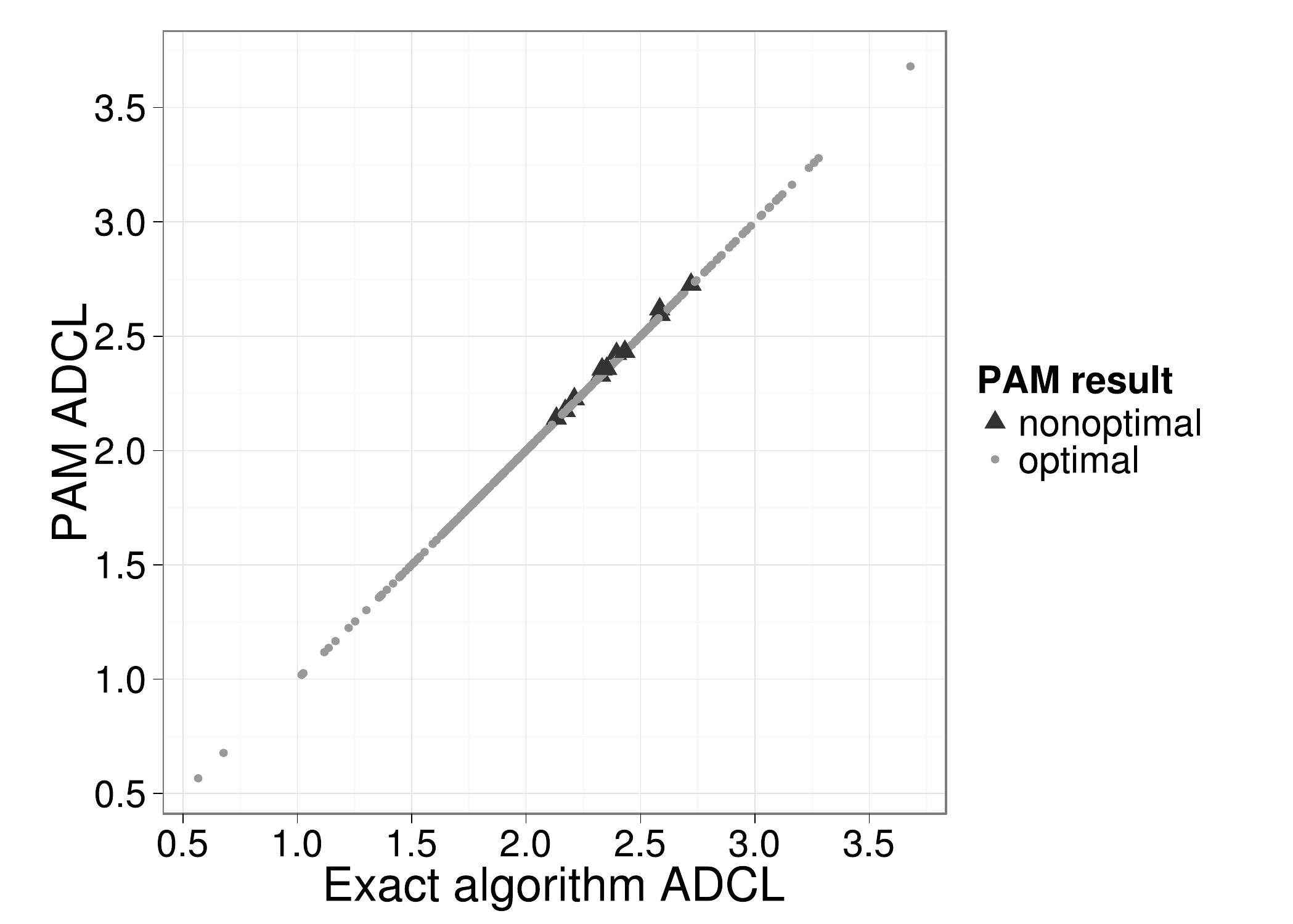}}
\end{center}
\caption{\
  Comparison between the ADCL results obtained by the exact algorithm and the PAM heuristic on the ``mass'' test set.
  The points are black triangles when the difference in ADCL between PAM and the exact algorithm is greater than $10^{-5}$.
}
\label{FIGpamOptimality}
\end{figure}
}
\newcommand{\refFIGpamOptimality}{5}

\newcommand{\FIGkTimeComplexity}{\
\begin{figure}[ht]
\begin{center}
  \forarxiv{\includegraphics[height=6cm]{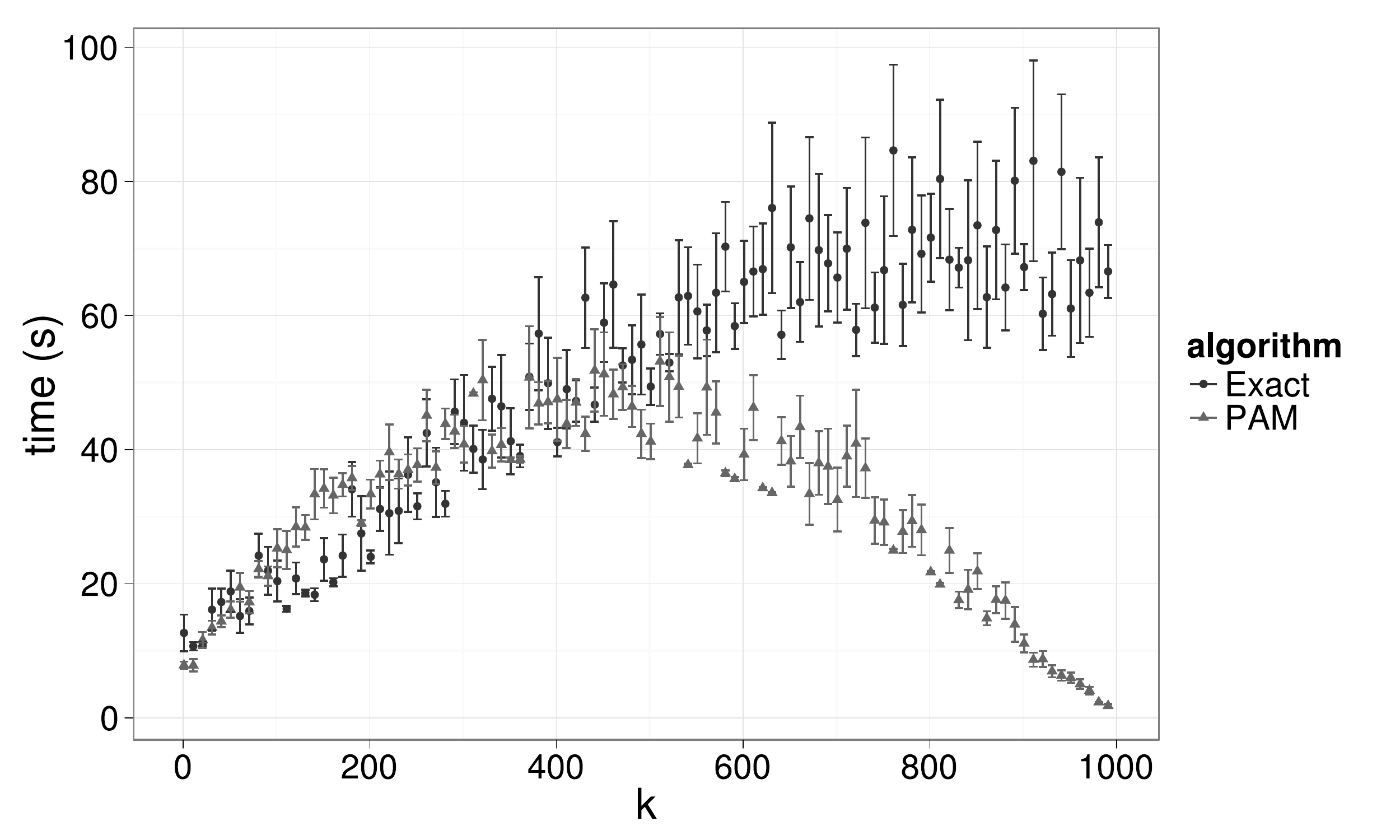}}
\end{center}
\caption{\
  Comparison of the time required to run the exact algorithm versus PAM with respect to the number of leaves selected to keep.
  Five trees were generated of 1,000 leaves each, and for each number $k$ from 1 to 991 congruent to 1 mod 10, both algorithms were run to keep $k$ of the leaves.
  The mean and standard errors are shown here.
}
\label{FIGkTimeComplexity}
\end{figure}
}
\newcommand{\refFIGkTimeComplexity}{6}

\newcommand{\FIGnTimeComplexity}{\
\begin{figure}[ht]
\begin{center}
  \forarxiv{\includegraphics[height=6cm]{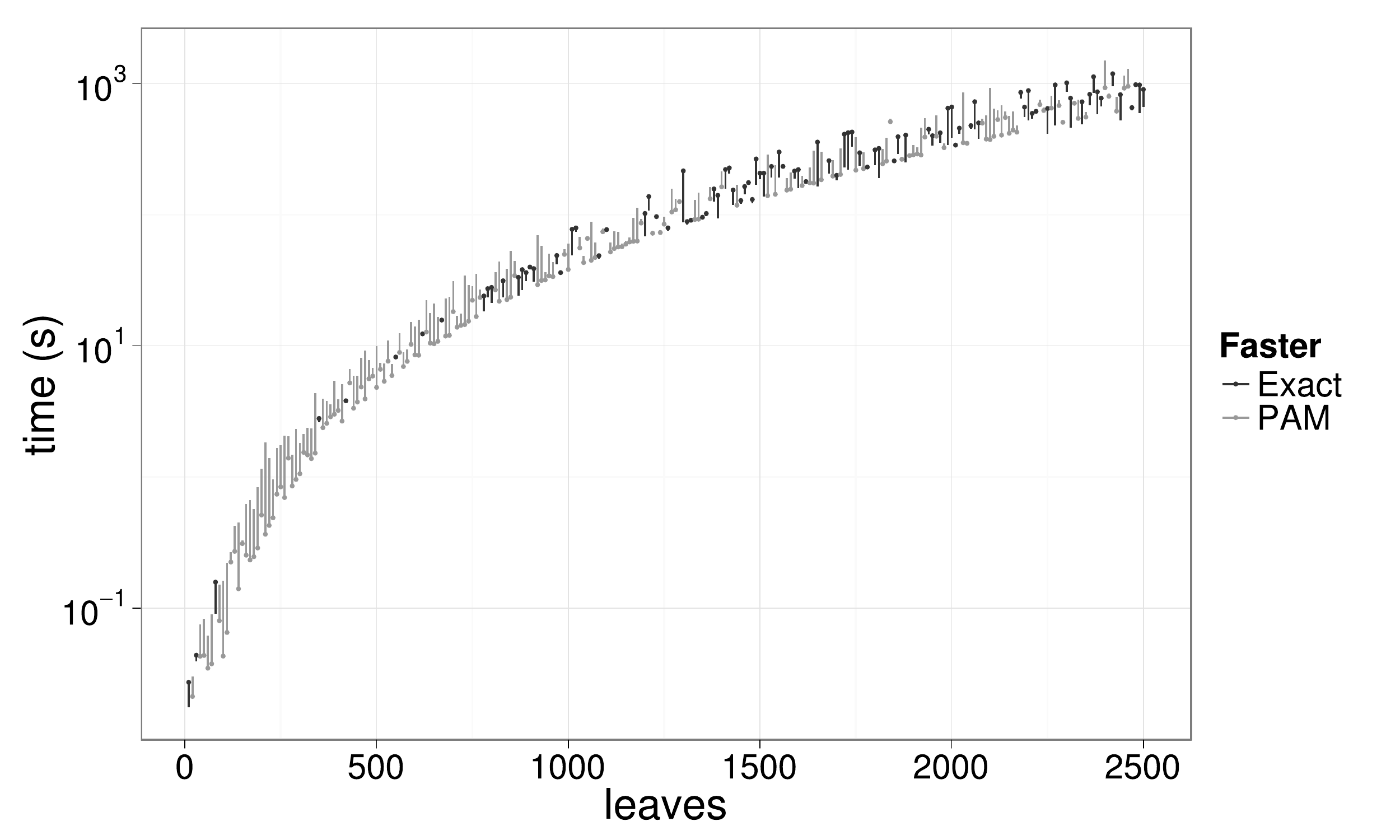}}
\end{center}
\caption{\
  Comparison of the time required to run the exact algorithm versus PAM with respect to the number of leaves in the original tree.
Trees were generated with 10 to 2,500 leaves in increments of 10 leaves; each tree was pruned to half of the original number of leaves.
  The result of running each algorithm on a given tree is shown here as a single line with $x$-position equal to the number of leaves of that tree, while the two $y$-positions of the line show the times taken by the two algorithms; when the exact algorithm was faster the line is black and when PAM was faster the line is gray.
  A point on each line shows the time for the PAM algorithm.
}
\label{FIGnTimeComplexity}
\end{figure}
}
\newcommand{\refFIGnTimeComplexity}{7}

\newcommand{\FIGchimera}{\
\begin{figure}[ht]
\begin{center}
  \forarxiv{\centerline{\includegraphics[height=6cm]{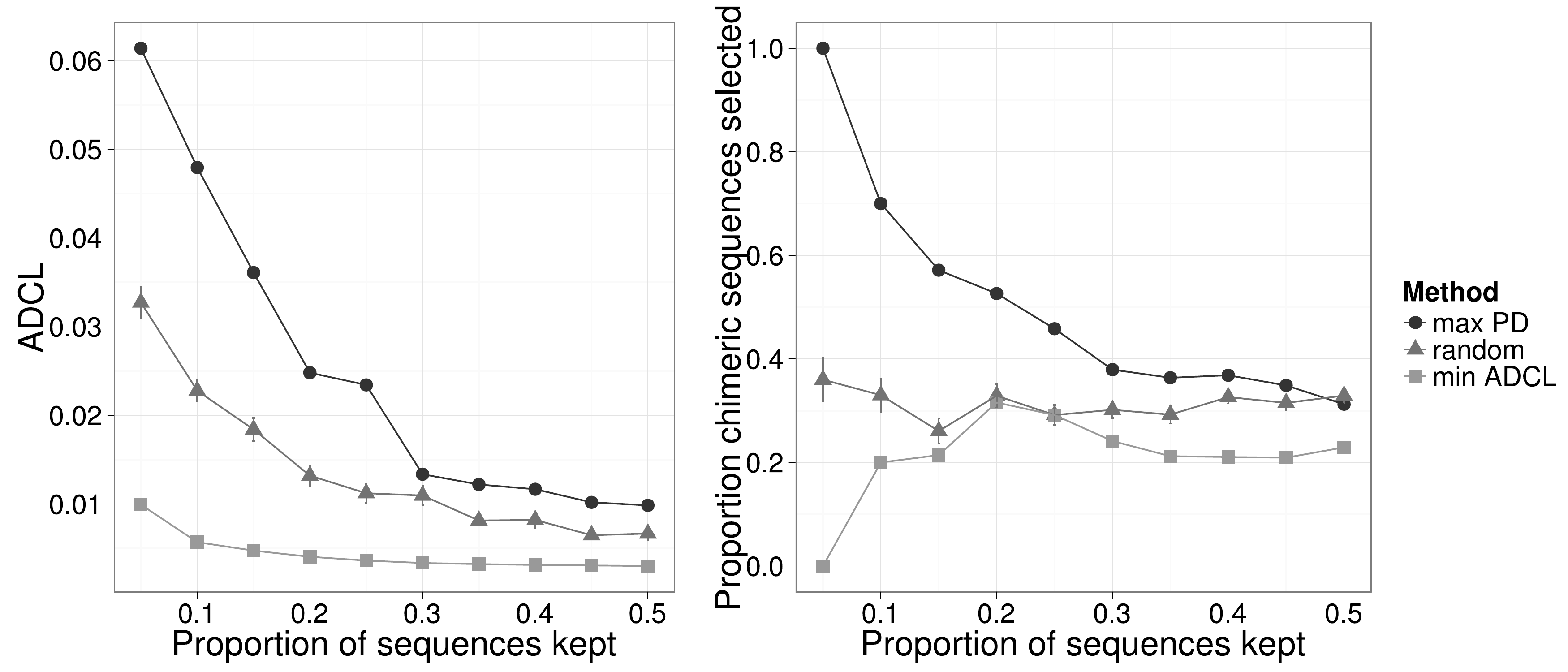}}}
\end{center}
\caption{\
  ADCL values and proportion of chimeric sequences kept for random selection, PD maximization, and ADCL minimization run on a set of Enterobacteriaceae 16s sequences along with chimeras from the same family identified with UCHIME.
}
\label{FIGchimera}
\end{figure}
}
\newcommand{\refFIGchimera}{8}

\newcommand{\TABlocalMinimum}{\
\begin{table}[ht]
\forarxiv{\
\begin{center}
\begin{tabular}{cc}
  \underline{subset} & \underline{ADCL} \\
  $\{n_0,n_1\}$ & 4.500 \\
  $\{n_0,n_2\}$ & 4.725 \\
  $\{n_0,n_3\}$ & 3.000 \\
  $\{n_1,n_2\}$ & 3.975 \\
  $\{n_1,n_3\}$ & 4.750 \\
  $\{n_2,n_3\}$ & 4.150 \\
\end{tabular}
\end{center}
}
\caption{ADCL values for all subsets of two leaves for the example in Figure~\refFIGlocalMinimum.\qquad\qquad\qquad\qquad}
\end{table}
}
\newcommand{\refTABlocalMinimum}{1}

\newcommand{\refFIGbootstrap}{S1}
\newcommand{\refFIGnMemoryComplexity}{S2}
\newcommand{\refFIGpamCostPerIter}{S3}
\newcommand{\refFIGtradPamCostPerIter}{S4}

\begin{document}

\notforarxiv{
\begin{flushright}
Version dated: \today
\end{flushright}
\bigskip
\noindent RH: MINIMIZING THE AVERAGE DISTANCE TO A CLOSEST LEAF
\bigskip
\medskip
\begin{center}

\noindent{\Large \bf Minimizing the average distance to a closest leaf in a phylogenetic tree}
\bigskip

\noindent {\normalsize \sc
Frederick A Matsen IV$^1$,
Aaron Gallagher$^1$,
Connor McCoy$^1$}\\
\noindent {\small \it
$^1$
Program in Computational Biology, Fred Hutchinson Cancer Research Center, Seattle, WA, 91802, USA}\\
\end{center}
\medskip
\noindent{\bf Corresponding author:} Frederick A Matsen, Program in Computational Biology, Fred Hutchinson Cancer Research Center, Seattle, WA, 91802, USA; E-mail: matsen@fhcrc.org.\\
\vspace{1in}
}

\forarxiv{\
\title[Minimizing the average distance to a closest leaf]{Minimizing the average distance to a closest leaf in a phylogenetic tree}
\author{Frederick A Matsen IV}
\author{Aaron Gallagher}
\author{Connor McCoy}
\date{\today}
\begin{abstract}
}
\notforarxiv{
\subsubsection{Abstract}
}
  When performing an analysis on a collection of molecular sequences, it can be convenient to reduce the number of sequences under consideration while maintaining some characteristic of a larger collection of sequences.
  For example, one may wish to select a subset of high-quality sequences that represent the diversity of a larger collection of sequences.
  One may also wish to specialize a large database of characterized ``reference sequences'' to a smaller subset that is as close as possible on average to a collection of ``query sequences'' of interest.
  Such a representative subset can be useful whenever one wishes to find a set of reference sequences that is appropriate to use for comparative analysis of environmentally-derived sequences, such as for selecting ``reference tree'' sequences for phylogenetic placement of metagenomic reads.
  In this paper we formalize these problems in terms of the minimization of the Average Distance to the Closest Leaf (ADCL) and investigate algorithms to perform the relevant minimization.
  We show that the greedy algorithm is not effective, show that a variant of the Partitioning Among Medoids (PAM) heuristic gets stuck in local minima, and develop an exact dynamic programming approach.
  Using this exact program we note that the performance of PAM appears to be good for simulated trees, and is faster than the exact algorithm for small trees.
  On the other hand, the exact program gives solutions for all numbers of leaves less than or equal to the given desired number of leaves, while PAM only gives a solution for the pre-specified number of leaves.
  Via application to real data, we show that the ADCL criterion chooses chimeric sequences less often than random subsets, while the maximization of phylogenetic diversity chooses them more often than random.
  These algorithms have been implemented in publicly available software.
\forarxiv{
\end{abstract}
\maketitle
\section{Introduction}
}

\notforarxiv{
\noindent (Keywords: sequence selection; mass transport; phylogenetic diversity)\\
\vspace{1.5in}
}

This paper introduces a method for selecting a subset of sequences of a given size from a pool of candidate sequences in order to solve one of two problems.
The first problem is to find a subset of a given collection of sequences that are representative of the diversity of that collection in some general sense.
The second is to find a set of ``reference'' sequences that are as close as possible on average to a collection of ``query'' sequences.

Algorithms for the first problem, selecting a diverse subset of sequences from a pool based on a phylogenetic criterion, have a long history.
The most well used such criterion is maximization of \emph{phylogenetic diversity} (PD), the total branch length spanned by a subset of the leaves \citep{faith1992conservation}.
The most commonly cited applications of these methods is to either select species to preserve \citep{faith1992conservation}, or to expend resources to perform sequencing \citep{pardi2005species,wu2009phylogeny}.
It is also commonly used for selecting sequences that are to be used as a representative subset that span the diversity of a set of sequences.

PD is a useful objective function that can be maximized very efficiently \citep{minh2006phylogenetic}, but it has some limitations when used for the selection of representative sequences.
Because maximizing the phylogenetic diversity function explicitly tries to choose sequences that are distant from one another, it tends to select sequences on long pendant branches \citep{bordewich2008selecting}; these sequences can be of low quality or otherwise different than the rest of the sequences.
Furthermore, PD has no notion of weighting sequences by abundance, and as such it can select artifactual sequences or other rare sequences that may not form part of the desired set of sequences.
This motivates the development of algorithms that strike a balance between centrality and diversity, such that one finds central sequences within a broad diversity of clusters.

The second problem addressed by the present work is motivated by modern genetics and genomics studies, where it is very common to learn about organisms by sequencing their genetic material.
When doing so, it is often necessary to find a collection of sequences of known origin with which to do comparative analysis.
Once these relevant ``reference'' sequences are in hand, a hypothesis or hypotheses on the unknown ``query'' sequences can be tested.

Although it is easy to pick out reference sequences that are close to an individual query sequence using sequence similarity searches such as BLAST, we are not aware of methods that attempt to find a collection of reference sequences that are close on average to a \emph{collection} of query sequences.
Such a method would have many applications in studies that use phylogenetics.
In phylogenomics, sequences of known function are used to infer the function of sequences of unknown function \citep{eisen1995evolution,eisen1998phylogenomics,engelhardt2005protein}.
In the study of HIV infections, hypotheses about the history of infection events can be phrased in terms of clade structure in phylogenetic trees built from both query and reference sequences \citep{piantadosi2007chronic}.
In metagenomics, it is now common to ``place'' a read of unknown origin into a previously constructed phylogeny \citep{berger2011performance,matsen2010pplacer}.
Each of these settings requires a set of reference sequences that are close on average to the collection of query sequences.

One approach to picking reference sequences would be to pick every potentially relevant sequence, such as all HIV reference sequences of the relevant subtype, but this strategy is not always practical.
Although many strategies in phylogenetics have been developed to speed inference, most analyses still require quadratic or greater execution time.
On the other hand, the number of sequences available to do comparative analysis is growing at an exponential pace.
This motivates strategies to pick useful subsets of sequences.

It may seem ironic that in order to find a useful subset of sequences for phylogenetics, we propose a fairly complex algorithm to use on a tree that has already been built; there are several reasons that have led us to develop this methodology.
First, tree-building methods vary widely in their running time, from sub-quadratic time methods \citep{price2010fasttree} to very computationally expensive Bayesian methods that model many aspects of the mutation process.
Similarly, analytic methods taking a tree as input may scale poorly with large numbers of taxa.
When a dataset is too large for an expensive method, our algorithm can be used in conjunction with a fast/approximate phylogenetic method to pick a subset of sequences to use in the more complex method.
Second, we note that there is a remarkable quantity of sequences for certain loci, such as over 2 million 16s sequences from Release 10 of the RDP database \citep{cole2009ribosomal}.
Because this number will continue to increase, and many of these sequences are redundant, we feel the need to have a principled method useful for curators to pick sequences that can form a representative subset of these large databases.
Others can then use the results of this curation process without having to run the algorithm themselves.

The objective is simple: select the collection of sequences that are on average as close as possible in terms of phylogenetic relatedness to the set of input sequences.
We now more formally state the two problems described above (Fig.~\refFIGmassTransport).
\begin{prob}
  \label{prob:vorotree}
  For a given phylogenetic tree $T$ and desired number of leaves $k$, find a $k$-element subset $X$ of the leaves $L$ that minimizes the Average Distance from each leaf in $L$ to its Closest Leaf (ADCL) in $X$.
\end{prob}

We emphasize that the distance is calculated between each leaf and its closest representative in $X$.

\begin{prob}
  \label{prob:voronoi}
  Given $T$ and $k$ as before, but let $R \subset L$ be a set of ``reference sequences''.
  Find the $k$-element subset $X$ of $R$ that minimizes the Average Distance of the leaves in $L \setminus R$ (the ``query sequences'') to their Closest Leaf in $X$.
\end{prob}
Recalling that the branch length between two sequences is typically the expected number of substitutions per site for those sequences, we are usually calculating the average expected number of substitutions relating each sequence to its closest selected leaf.

These criteria, along with generalizations, can be expressed in a single framework in terms of ``mass transport'' \citep{villani2003topics} on trees as follows (Fig.~\refFIGmassTransport).
In this framework, the ``work'' needed to move a mass $m$ a distance $d$ is defined to be $m$ times $d$.
For Problem~\ref{prob:vorotree} above, assume we are interested in selecting sequences according to the first criterion on a tree with $n$ sequences.
Distribute mass $1/n$ at each of the leaves, and then find the set $X$ of $k$ leaves such that the work required to move the mass to one of the leaves in $X$ is minimized.
This is equivalent to minimizing the ADCL criterion, because the optimal solution will have all of the mass for a single leaf being transported to its closest included leaf, incurring a cost of that distance divided by $n$; the sum of these individual quantities of work will be equal to the ADCL.

\forarxiv{\FIGmassTransport}

In a similar way, the second criterion can be phrased as evenly dividing a unit of mass among the tips of $L \setminus R$, and finding a set $X$ of leaves in $R$ minimizing work as before.
In this second criterion, call the tree induced by the reference sequences $R$ the ``reference tree''.
Because mass can only be transported to reference tree leaves and not query leaves, all of the mass of the query sequences must first be transported somewhere on the reference tree.
This amount of work is a fixed cost, and thus we can just think of the mass for a subtree composed of only query sequences as appearing at the attachment location of that query-only subtree to the reference tree.
This change of perspective will change the magnitude of, but not the differences between, the ADCL values, thus giving an equivalent solution to Problem~\ref{prob:voronoi}.

A further motivation for considering mass at internal nodes of a tree comes from \emph{phylogenetic placement}, i.e.\ the mapping of query sequences into a tree built from the reference sequences.
This collection of placements can then be thought of as a collection of mass on the tree, and the optimization can proceed as above.
The transition of placements to mass distribution can include ``spreading'' out mass according to uncertainty in placement \citep{matsen2010pplacer,EvansMatsenPhyloKR12}.
Because of the speed of placement algorithms, this can be a useful way of proceeding when the set of query sequences is large.
We have previously used mass transport to measure the differences between collections of placements \citep{EvansMatsenPhyloKR12}.
In this context, a collection of query sequences can be mapped onto the tree and used to pick an optimal subset of reference sequences.

Arbitrary distributions of mass on the tree are also possible.
These distributions may arise from transforms of placement distributions.
Alternatively, they may arise by assigning an arbitrary value to various locations on the tree; these values may convey the importance of regions of a tree for some analysis.

Because all of these can be formulated in terms of mass on a phylogenetic tree, rather than considering Problems~\ref{prob:vorotree} and~\ref{prob:voronoi} separately, we solve the following generalization of both of them:
\begin{prob}
  \label{prob:masstrans}
  Given a mass distribution $\mu$ on a phylogenetic tree $T$ with $n$ leaves and some $0 < k < n$, find the $k$-element subset $X$ of the leaves of $T$ such that the work required to move the mass $\mu(x)$ at point $x \in T$ to $x$'s closest leaf in $X$ is minimized across all $k$-element subsets of the leaves of $T$.
\end{prob}
We will still call this problem ``minimizing ADCL'' because it attempts to minimize the average distance to the closest leaf, where now the average is weighted by the mass distribution.
It should also be pointed out that the distances used in the ADCL framework are the distances in the original tree.
When leaves are pruned out of the tree, a tree built \emph{de novo} on this reduced set will have different branch lengths than the original tree with branches pruned out; we do not attempt to correct for that effect here.
A more formal statement of Problem~\ref{prob:masstrans} is made in the Appendix.

Problem~\ref{prob:vorotree} is equivalent to the DC1 criteria independently described in chapter 5 of Barbara Holland's Ph.D. thesis \citep{holland2001thesis}.
She writes out the criterion (among others), discusses why it might be biologically relevant, describes the computational complexity of the brute-force algorithm, and does some experiments comparing the brute-force to the greedy algorithm.
She also describes $L^2$ and $L^\infty$ versions of Problem~\ref{prob:vorotree}.

We also note that the work described here shares some similarities with the Maximizing Minimum Distance (MMD) criterion of \citet{bordewich2008selecting}.
In that criterion, the idea is to select the subset $X$ of leaves such that the minimum distance between any two leaves in $X$ is maximized across subsets of size $k$.
The MMD criterion has more similarities with PD maximization than it does with Problem~\ref{prob:vorotree}. Moreover, the MMD analog of Problem~\ref{prob:voronoi} (minimizing the maximum distance of a reference sequence to a query sequence) would be highly susceptible to off-target query sequences, such as sequences that are similar to but not actually homologous to the reference set.
Because of this difference in objective functions, we have not attempted a comparison with MMD here.

The analogous problem in the general non-phylogenetic setting is the classical $k$-medoids problem where $k$ ``centers'' are found minimizing the average distance from each point to its closest center.
The PAM algorithm (described below) is a general heuristic for such problems, although our exact algorithm, which is based on additivity of distances in a tree, will not work.
It appears that the complexity of exact $k$-medoids is not known and is only bounded above by the obvious brute-force bound.
The simpler setting of $k$-means in the plane has been shown to be NP-hard by \citet{mahajan2009planar}.

We emphasize that for the purposes of this paper, we assume that $k$ has been chosen ahead of time.
Although we consider choosing an appropriate $k$ to be an interesting direction for future work, as described in the discussion, the choice will depend substantially on the goals of the user.

\section{Methods}

In this section we will investigate methods to minimize ADCL for a given mass distribution as in Problem~\ref{prob:masstrans}.
We will first show that the greedy algorithm fails to provide an optimal solution, then describe a variant of the Partitioning Around Medoids algorithm that finds a local minimum of ADCL, and then describe our dynamic program that is guaranteed to find a global minimum.

For all of these algorithms, the structure of the tree is not re-estimated and branch lengths not re-optimized after removal of leaves.
Indeed, the tree is not changed at all, rather the removal is from the set of selected leaves.

\subsection{Optimization via greedy leaf pruning}
The PD minimization problem is known to be solved exactly by a greedy leaf pruning algorithm \citep{steel2005phylogenetic}, and by analogy a reasonable first attempt might be to try to apply the same approach.
\begin{alg}[Greedy leaf pruning]
  Given a tree $T$, a mass distribution $\mu$, and a desired number of leaves $\maxleaves$, start with $X$ being all of the leaves of the tree.
  \begin{enumerate}
    \item If $|X| = \maxleaves$ then stop.
    \item Find the $\ell$ minimizing the ADCL of $X \setminus \{\ell\}$
    \item Remove $\ell$ from $X$ and return to (1).
  \end{enumerate}
  \label{alg:greedy}
\end{alg}
A similar algorithm, which instead greedily \emph{adds} sequences to the chosen set, was independently described by Barbara Holland in her thesis \citep{holland2001thesis}.

\begin{prop}
  Algorithm~\ref{alg:greedy} does not find an optimal solution to Problem~\ref{prob:masstrans}.
\end{prop}

\begin{proof}
  Fix a three-taxon tree with a single internal node, and label the leaves $n_0$, $n_1$, and $n_2$.
  Assign mass $m$ to leaves $n_0$ and $n_1$, and assign mass $m-\epsilon$ to leaf $n_2$.
  Let the edges going to $n_0$ and $n_1$ have length $x$, and the edge going to $n_2$ have length $y$.
  Choose these values satisfying $0 < \epsilon < m$ and $0 < y < x$, and such that $\epsilon \cdot (x+y) < m \cdot (x - y)$.

  A greedy algorithm will delete leaf $n_2$ as a first step, because deleting either $n_0$ or $n_1$ increases ADCL by $m \cdot (x+y)$, while deleting $n_2$ increases it by $(m - \epsilon) \cdot (x+y)$.
  However, deleting $n_0$ and $n_1$ at once leads to an ADCL of $2m \cdot (x+y)$, while the other options give an ADCL of $m \cdot 2x + (m-\epsilon) \cdot (x+y)$.
  By our choice of $m, \epsilon, x$, and $y$,
  \[
    2m \cdot (x+y) < m \cdot 2x + (m-\epsilon) \cdot (x+y).
  \]
  Therefore removing $n_0$ and $n_1$ is optimal, while the greedy algorithm removes $n_2$ in its first step.
\end{proof}
This approach has shown poor enough performance in practice compared to the one in the next section (results not shown) that we have not pursued efficient optimization.

\subsection{Optimization via Partitioning Among Medoids}
A different way to minimize ADCL is to adapt heuristic algorithms for the so-called $k$-medoids problem.
The objective of $k$-medoid clustering is the same as $k$-means clustering, except that the cluster centers $X$ must be chosen to be elements of the set $L$ being clustered.
Those chosen centers $X \subset L$ are called ``medoids.''
That is, the objective is to find a subset $X$ of elements minimizing the average distance from each element $y$ of $L$ to the closest element $x_y$ of $X$.
Problem~\ref{prob:vorotree} can be expressed as a standard $k$-medoid problem, where the points are leaves of the tree and the distances between them are distances between those leaves of the tree.
One common approach for the $k$-medoid problem is called the Partitioning Around Medoids (PAM) algorithm \citep{patternRecognitionBook}.
This algorithm starts with a random selection of $k$ medoids, then executes a hill-climbing heuristic to improve the relevant objective function.

Problem~\ref{prob:voronoi} can also be formulated as a variant of PAM that we now describe.
The same algorithm can be used for Problem~\ref{prob:masstrans} in the (common) case of a discrete mass distribution.
Let $J$ be an arbitrary set of items.
Assume we are given a distance matrix $M$ that measures the distance from a set of ``leaves'' $L$ to the items in $J$, as well as some $0 < k \le |L|$.
The goal is to find a $k$-element set of leaves $X \subset L$ minimizing the objective function, which is the average distance of each item in $J$ to its closest leaf in $X$.
\begin{alg}[PAM variant]
  Initialize $X \subset L$ as a random selection of $k$ leaves.
  Repeat the following process until an iteration over every value in $X$ does not strictly decrease the objective function.
  \begin{enumerate}
    \item For a single $i$ in $X$; remove it from $X$ and try adding every other $j \in L \setminus X$ to $X$ in its place.
    \item Keep the best such exchange if it decreases the objective function.
    \item Continue with the next $i$ in $X$.
  \end{enumerate}
  \label{alg:pam}
\end{alg}
This differs from the traditional formulation of PAM in two ways.
First, the set $J$ is not necessarily identical to $L$.
Second, whereas PAM examines every combination from $X \times (L \setminus X)$, choosing the exchange that most decreases the objective function, this variant only examines the potential exchanges for a single medoid at a time, making the exchange of these that most decreases the objective function before continuing with the next medoid.

The complexity of this PAM variant is $O(k (|X| - k) |J|)$ for every iteration.

\forarxiv{\FIGlocalMinimum}

PAM will always find a local minimum in terms of these pairwise exchanges; however it will not always find a global minimum as shown in Figure~\refFIGlocalMinimum\ and Table~\refTABlocalMinimum.
In that example, there are four masses $\{a, b, c, d\}$ on a four taxon tree.
Assume we are trying to find the pair of leaves minimizing ADCL, and PAM selects $\{n_1,n_2\}$ as a starting set.
The optimal solution is to take $\{n_0,n_3\}$.
Because it only swaps a single pair of sequences, and the ADCL increases for every such swap from $\{n_1,n_2\}$, it will not be able to escape this local minimum (Table~\refTABlocalMinimum).

\forarxiv{\TABlocalMinimum}

\subsection{Exact algorithm}

We also present the following exact dynamic program.
We note that the exact algorithm gives solutions for all numbers of leaves less than or equal to the given desired number of leaves $k$, which can be useful when the best $k$ is to be inferred from the data.
This section will give a high-level overview of the exact algorithm; for a complete description see the Appendix.

Our exact algorithm is a dynamic program that proceeds from a chosen ``root'' out to the leaves then back to the root of the tree.
Assume the dynamic program has descended into a subtree $S$ of $T$.
The optimal solution will allow some number of leaves to be used within $S$, and will have some amount, direction, and distance of mass transport through the root of $S$.
However, by the nature of a dynamic program, this number of leaves and mass transportation characteristics are necessarily not known when the algorithm is visiting the subtree $S$.
For this reason, the algorithm builds a collection of ``partial solutions'' for every subtree and number of selected leaves less than or equal to $k$; these partial solutions are indexed by the amount and direction of mass transport going through the root of $S$ and only specify where the mass within that subtree will go (Fig.~\refFIGrmdRmp).
When progressing proximally (towards the root) through an internal node that is the root of a subtree $S$, the candidate list of partial solutions for $S$ is built from every combination of partial solutions for the subtrees of $S$.
Because this combination is done for every allocation of leaves to subtrees of total number of leaves less than or equal to $k$, the final output of this algorithm is solutions for every number of allowed leaves less than or equal to $k$.

\forarxiv{\FIGrmdRmp}

These solution sets can become very large, and it is necessary to cut them down in order to have a practical algorithm.
Ideally, this dynamic program would only maintain partial solutions that could be optimal for some number of leaves and some amount and direction of mass transport through the root of $S$.
In fact it is possible to only keep \emph{exactly} that set of partial solutions using methods from geometry.
The partial solutions can be partitioned by the number of leaves used and the direction of mass transport through the root of $S$.
These solutions will either be \emph{root mass distal} (RMD) solutions, where the solution may accept mass to flow into the subtree from the outside, or \emph{root mass proximal} (RMP) solutions, where the solution sends mass through the root of the subtree to the outside.
Note that the distinction between these two types of partial solutions concerns the flow of the mass through the root of the subtree only.
For example, RMP solutions with a non-empty leaf set have some mass flowing towards those leaves and a (possibly empty) amount of mass flowing proximally out of the root.
Whether a partial solution is an RMP or an RMD solution specifies its \emph{mass direction class}.

Thus for the dynamic program we will solve the optimal mass transport for all possible contexts of the rest of the tree: when the partial solution has \emph{proximal mass} going proximally through the root of $S$ to some leaf of unknown distance in the rest of the tree, or some unknown amount of mass descending to a leaf in $S$.
For an RMP solution, the amount of work required in a given partial solution to move the mass in $S$ to some selected leaf is equal to the mass transport within $S$ plus the amount of work required to move the proximal mass of that partial solution some distance $x$ away from the root of $S$.
The amount of work is linear in $x$, with $y$-intercept the amount of work within $S$, and with slope equal to the amount of mass that moves outside of $S$.
For a given partial solution, we can plot the contribution of the mass in $S$ to the ADCL as a function of $x$.
In a similar way, we can plot the amount of work required for an RMD solution, except this time the appropriate parameter $x$ to use is the amount of mass that comes distally (away from the root) through the root of $S$.
The work is again linear in $x$, and the $y$-intercept is again the amount of work within $S$, but with slope equal to the distance to the closest leaf that is selected in $S$.
These lines will be called \emph{subwork} lines, and the parameter $x$ will be called the \emph{subwork parameter}.

\forarxiv{\FIGlowerHull}

The only partial solutions that could form part of an optimal solution are those that are optimal for some value of the subwork parameter (Fig.~\refFIGlowerHull).
This optimization can be done using well-known algorithms in geometry.
Imagine that instead of considering the minimum of a collection of subwork lines, we are using these lines to describe a subset of the plane using inequalities.
Some of these inequalities are spurious and can be thrown away without changing the subset of the plane; the rest are called \emph{facets} \citep{ziegler1995lectures}.
Our implementation uses Komei Fukuda's \cddlib\ implementation \citep{cddlib} of the double description method \citep{fukuda1996double} to find facets.
The complexity of our exact algorithm is difficult to assess, given that in the words of \citet{fukuda1996double}, ``we can hardly state any interesting theorems on [the double description algorithm's] time and space complexities.''
We note that our code uses the floating point, rather than exact arithmetic, version of \cddlib\ because branch lengths are typically obtained by numerical optimization to a certain precision.
For this reason, our implementation is susceptible to rounding errors, and ADCLs are only compared to within a certain precision in the implementation.
We note that a solution that is optimal when restricted to a subproblem need not be optimal itself.

\section{Results}

We have developed and implemented algorithms to minimize the Average Distance to the Closest Leaf (ADCL) among subsets of leaves of the reference tree.
These algorithms are implemented as the \minadcltree\ (Problem~\ref{prob:vorotree}) and \minadcl\ (Problems~\ref{prob:vorotree},~\ref{prob:masstrans}) subcommands of the \rppr\ binary that is part of the \pplacer\ (\url{http://matsen.fhcrc.org/pplacer/}) suite of programs.
The code for all of the \pplacer\ suite is freely available (\url{http://github.com/matsen/pplacer}).

Our PAM implementation follows Algorithm~\ref{alg:pam}.
We found this variant, which makes the best exchange at each medoid rather than the best exchange over all medoids, to converge two orders of magnitude more rapidly than a traditional PAM implementation (Figs.~\refFIGpamCostPerIter, \refFIGtradPamCostPerIter).

We used simulation to understand the frequency with which PAM local minima are not global minima as in Figure~\refFIGlocalMinimum, as well as the relative speed of PAM and the exact algorithm.
For tree simulation, random trees were generated according to the Yule process \citep{yule1925mathematical}; branch lengths were drawn from the gamma distribution with shape parameter 2 and scale parameter 1.
We evaluated three data sets: two ``tree'' sets and one ``mass'' set.
For the ``tree'' test sets, trees were randomly generated as described, resulting in 5 trees of 1,000 leaves each, and a collection of trees with 10 to 2,500 leaves in increments of 10 leaves.
Problem~\ref{prob:vorotree} was then solved using each algorithm for each ``tree'' test set; for the 1,000 leaf trees $k$ was set to each number from 1 to 991 congruent to 1 mod 10, and for the large collection of trees $k$ was set to half the number of leaves.
For the ``mass'' test set, one tree was built for each number of leaves from 5 to 55 (inclusive); for each of these trees, $m$ masses were assigned to uniformly selected edges of the tree at uniform positions on the edges, where $m$ ranged from 5 to 95 masses in 10-mass increments.
All of these simulated test sets have been deposited in Dryad (\url{http://datadryad.org/handle/10255/dryad.41611}).
Problem~\ref{prob:voronoi} was then solved using each algorithm with $k$ equal to (the ceiling of) half the number of leaves.

\forarxiv{\FIGpamOptimality}

The PAM heuristic typically works well.
Although we have shown above that the PAM algorithm does get stuck in local minima (Fig.~\refFIGlocalMinimum), it did so rarely on the ``mass'' data set (Fig.~\refFIGpamOptimality); similar results were obtained for the ``tree'' data set (results not shown).
As might be expected, PAM displays the greatest speed advantage in when $k$ is rather large on the ``tree'' data set (Fig.~\refFIGkTimeComplexity).
PAM is slowest for $k$ equal to $n/2$ because that value of $k$ has the largest number of possible $k$-subsets; once $k > n/2$ it gets faster because there are fewer choices as far as what to select.
PAM is faster for small trees than the exact algorithm on the ``tree'' data set (Fig.~\refFIGnTimeComplexity), and uses less memory (Fig.~\refFIGnMemoryComplexity).
We note in passing that the ADCL improvement from PAM is not monotonically non-increasing, which is to say that it is possible to have a small improvement followed by a large improvement.

\forarxiv{\FIGkTimeComplexity}

\forarxiv{\FIGnTimeComplexity}

The currently available tools for automatic selection of reference sequences include the use of an algorithm that maximizes PD or using a random set of sequences \citep{redd2011identification}.
Others have pointed out that long pendant branch lengths are preferentially chosen by PD maximization, even when additional ``real'' diversity is available \citep{bordewich2008selecting}.
Long pendant branch lengths can be indicative of problematic sequences, such as chimeras or sequencing error.

We designed an experiment to measure the extent to which the ADCL algorithm would pick problematic sequences compared to PD maximization and random selection.
We downloaded sequences with taxonomic annotations from Genbank belonging to the family Enterobacteriaceae, and identified chimeric sequences using UCHIME \citep{edgar2011uchime}.
We built a tree of consisting of these sequences and other sequences from the same species from the RDP database release 10, update 28 \citep{cole2009ribosomal}.
Five sequences were chosen for each species (when such a set existed).
Remaining RDP sequences from these species were then placed on this tree using \pplacer\ \citep{matsen2010pplacer} and the full algorithm was used to pick some fraction $x$ of the reference sequences.
In this case, using the full algorithm to minimize ADCL was less likely to choose chimeric sequences than choosing at random, while PD maximization was more likely to choose chimeric sequences (Fig.~\refFIGchimera).
The ADCL for the full algorithm was substantially lower than for either PD maximization or random subset selection, as would be expected given that ADCL is explicitly minimized in our algorithms.

We are not proposing this algorithm as a new way to find sequences of poor quality, rather, it is a way of picking sequences that are representative of the local diversity in the tree.
The chimera work above was to make the point that artifactual sequences clearly not representative of actual diversity do not get chosen, while they do using the PD criterion.
We also note that because bootstrap resampling can change branch length and tree topology, the minimum ADCL set is not guaranteed to be stable under bootstrapping.
However, in the cases we have evaluated, the minimum ADCL value itself is relatively stable to bootstrap resampling when $k$ is not too small (Fig.~\refFIGbootstrap).

\forarxiv{\FIGchimera}

\section{Discussion}

In this paper we described a simple new criterion, minimizing the Average Distance to the Closest Leaf (ADCL), for finding a subset of sequences that either represent the diversity of the sequences in a sample, or are close on average to a set of query sequences.
In doing so, abundance information is taken into account in an attempt to strike a balance between optimality and centrality in the tree.
In particular, this criterion is the only way of which we are aware to pick sequences that are phylogenetically close on average to a set of query sequences.
We have also investigated means of minimizing the ADCL, including a heuristic that performs well in practice and an exact dynamic program.
ADCL minimization appears to avoid picking chimeric sequences.

The current implementations are useful for moderate-size trees; improved algorithms will be needed for large-scale use (Fig.~\refFIGnTimeComplexity).
We have found present algorithms to be quite useful in a pipeline that clusters query sequences by pairwise distance first, then retrieves a collection of potential reference sequences per clustered group of query sequences, then uses the ADCL criterion for selecting potential reference sequences amongst those (manuscript under preparation).
This has the advantage of keeping the number of input sequences within a manageable range, as well as ensuring that the number of reference sequences is comprehensive across the tree.

The computational complexity class of the ADCL optimization problem is not yet clear.

Because of the special geometric structure of the problem, there is almost certainly room for improvement in the algorithms used to optimize ADCL\@.
Only a subset of the possible exchanges need to be tried in each step of the PAM algorithm, and more intelligent means could be used for deciding which mass needs to be reassigned, similar to the methods of \citet{zhang2005new}.
A better understanding of situations such as those illustrated by Figure~\refFIGlowerHull\ could lead to an understanding of when PAM becomes stuck in local optima.
The geometric structure of the optimality intervals could be better leveraged for a more efficient exact algorithm.
The PAM algorithm may also reach a near-optimal solution quickly, then use substantial time making minimal improvements to converge to the minimum ADCL (Fig.~\refFIGpamCostPerIter).
If an approximate solution is acceptable, alternate stopping criteria could be used.

In future work, we also plan on investigating the question of what $k$ is appropriate to use for a given phylogenetic tree given certain desirable characteristics of the cut-down set.
We note that using the exact algorithm makes it easy to find a $k$ that corresponds to an upper bound for ADCL, but the choice of an appropriate upper bound depends on the application and priorities of the user.
For example, in taxonomic assignment, some may require subspecies-level precision while others require assignments only at the genus or higher level; the needs for ADCL would be different in these different cases.

\section{Funding}
The authors are supported in part by NIH grant R01 HG005966-01 and by startup funds from the Fred Hutchinson Cancer Research Center.

\section{Acknowledgments}
FAM thanks Christian von Mering for reminding him that phylogenetic diversity often picks sequences of low quality, and Steve Evans for helping think about the formulation of mass on trees, as well as pointing out the ``medoids'' literature.
All authors are grateful to Sujatha Srinivasan, Martin Morgan, David Fredricks, and especially Noah Hoffman for helpful comments.
Aaron Darling, Barbara Holland, and two anonymous reviewers provided feedback that greatly improved the form and content of the manuscript.

\notforarxiv{\newpage}

\bibliography{voronoi}

\begin{thebibliography}{27}
\providecommand{\natexlab}[1]{#1}
\providecommand{\url}[1]{\texttt{#1}}
\expandafter\ifx\csname urlstyle\endcsname\relax
  \providecommand{\doi}[1]{doi: #1}\else
  \providecommand{\doi}{doi: \begingroup \urlstyle{rm}\Url}\fi

\bibitem[Berger et~al.(2011)Berger, Krompass, and
  Stamatakis]{berger2011performance}
S.A. Berger, D.~Krompass, and A.~Stamatakis.
\newblock Performance, accuracy, and web server for evolutionary placement of
  short sequence reads under maximum likelihood.
\newblock \emph{Syst. biol.}, 60\penalty0 (3):\penalty0 291--302, 2011.

\bibitem[Bordewich et~al.(2008)Bordewich, Rodrigo, and
  Semple]{bordewich2008selecting}
M.~Bordewich, A.G. Rodrigo, and C.~Semple.
\newblock Selecting taxa to save or sequence: desirable criteria and a greedy
  solution.
\newblock \emph{Syst. biol.}, 57\penalty0 (6):\penalty0 825--834, 2008.

\bibitem[Cole et~al.(2009)Cole, Wang, Cardenas, Fish, Chai, Farris,
  Kulam-Syed-Mohideen, McGarrell, Marsh, Garrity, et~al.]{cole2009ribosomal}
JR~Cole, Q.~Wang, E.~Cardenas, J.~Fish, B.~Chai, RJ~Farris,
  AS~Kulam-Syed-Mohideen, DM~McGarrell, T.~Marsh, GM~Garrity, et~al.
\newblock The {Ribosomal} {Database} {Project}: improved alignments and new
  tools for {rRNA} analysis.
\newblock \emph{Nucleic acids res.}, 37\penalty0 (suppl 1):\penalty0
  D141--D145, 2009.

\bibitem[Edgar et~al.(2011)Edgar, Haas, Clemente, Quince, and
  Knight]{edgar2011uchime}
R.C. Edgar, B.J. Haas, J.C. Clemente, C.~Quince, and R.~Knight.
\newblock Uchime improves sensitivity and speed of chimera detection.
\newblock \emph{Bioinformatics}, 27\penalty0 (16):\penalty0 2194--2200, 2011.

\bibitem[Eisen(1998)]{eisen1998phylogenomics}
J.A. Eisen.
\newblock Phylogenomics: improving functional predictions for uncharacterized
  genes by evolutionary analysis.
\newblock \emph{Genome res.}, 8\penalty0 (3):\penalty0 163--167, 1998.

\bibitem[Eisen et~al.(1995)Eisen, Sweder, and Hanawalt]{eisen1995evolution}
J.A. Eisen, K.S. Sweder, and P.C. Hanawalt.
\newblock Evolution of the {SNF2} family of proteins: subfamilies with distinct
  sequences and functions.
\newblock \emph{Nucleic acids res.}, 23\penalty0 (14):\penalty0 2715--2723,
  1995.

\bibitem[Engelhardt et~al.(2005)Engelhardt, Jordan, Muratore, and
  Brenner]{engelhardt2005protein}
B.E. Engelhardt, M.I. Jordan, K.E. Muratore, and S.E. Brenner.
\newblock Protein molecular function prediction by {Bayesian} phylogenomics.
\newblock \emph{PLOS comp. biol.}, 1\penalty0 (5):\penalty0 e45, 2005.

\bibitem[Evans and Matsen(2012)]{EvansMatsenPhyloKR12}
S~N Evans and F~A Matsen.
\newblock {The phylogenetic Kantorovich-Rubinstein metric for environmental
  sequence samples}.
\newblock \emph{J. Royal Stat. Soc. (B)}, 74\penalty0 (3):\penalty0 569--592,
  2012.

\bibitem[Faith(1992)]{faith1992conservation}
D.P. Faith.
\newblock Conservation evaluation and phylogenetic diversity.
\newblock \emph{Biol. Conserv.}, 61\penalty0 (1):\penalty0 1--10, 1992.

\bibitem[Fukuda(2012)]{cddlib}
K.~Fukuda.
\newblock cddlib library.
\newblock \url{http://www.cs.mcgill.ca/~fukuda/soft/cdd_home/cdd.html}, 2012.
\newblock version 094f.

\bibitem[Fukuda and Prodon(1996)]{fukuda1996double}
K.~Fukuda and A.~Prodon.
\newblock Double description method revisited.
\newblock \emph{Combinat. comp. sci.}, pages 91--111, 1996.

\bibitem[Holland(2001)]{holland2001thesis}
B.~Holland.
\newblock \emph{Evolutionary analyses of large data sets: trees and beyond}.
\newblock PhD thesis, Massey University, 2001.
\newblock http://hdl.handle.net/10179/2078.

\bibitem[Mahajan et~al.(2009)Mahajan, Nimbhorkar, and
  Varadarajan]{mahajan2009planar}
M.~Mahajan, P.~Nimbhorkar, and K.~Varadarajan.
\newblock The planar k-means problem is {NP}-hard.
\newblock \emph{WALCOM: Algorithms and Computation}, pages 274--285, 2009.

\bibitem[Matsen et~al.(2010)Matsen, Kodner, and Armbrust]{matsen2010pplacer}
F.A. Matsen, R.B. Kodner, and E.V. Armbrust.
\newblock pplacer: linear time maximum-likelihood and {Bayesian} phylogenetic
  placement of sequences onto a fixed reference tree.
\newblock \emph{BMC bioinfo.}, 11\penalty0 (1):\penalty0 538, 2010.

\bibitem[Minh et~al.(2006)Minh, Klaere, and Von~Haeseler]{minh2006phylogenetic}
B.Q. Minh, S.~Klaere, and A.~Von~Haeseler.
\newblock Phylogenetic diversity within seconds.
\newblock \emph{Syst. biol.}, 55\penalty0 (5):\penalty0 769--773, 2006.

\bibitem[Motzkin et~al.(1983)Motzkin, Ralffa, Thompson, and
  Thrall]{motzkin1983double}
TS~Motzkin, H.~Ralffa, G.L. Thompson, and RM~Thrall.
\newblock The double description method.
\newblock In \emph{Theodore S. Motzkin: selected papers}, page~81. Birkhauser,
  1983.

\bibitem[Pardi and Goldman(2005)]{pardi2005species}
F.~Pardi and N.~Goldman.
\newblock Species choice for comparative genomics: Being greedy works.
\newblock \emph{PLOS genet.}, 1\penalty0 (6):\penalty0 e71, 2005.

\bibitem[Piantadosi et~al.(2007)Piantadosi, Chohan, Chohan, McClelland, and
  Overbaugh]{piantadosi2007chronic}
A.~Piantadosi, B.~Chohan, V.~Chohan, R.S. McClelland, and J.~Overbaugh.
\newblock Chronic {HIV}-1 infection frequently fails to protect against
  superinfection.
\newblock \emph{PLOS pathogens}, 3\penalty0 (11):\penalty0 e177, 2007.

\bibitem[Price et~al.(2010)Price, Dehal, and Arkin]{price2010fasttree}
M.N. Price, P.S. Dehal, and A.P. Arkin.
\newblock Fasttree 2--approximately maximum-likelihood trees for large
  alignments.
\newblock \emph{PLOS ONE}, 5\penalty0 (3):\penalty0 e9490, 2010.

\bibitem[Redd et~al.(2011)Redd, Collinson-Streng, Martens, Ricklefs, Mullis,
  Manucci, Tobian, Selig, Laeyendecker, Sewankambo,
  et~al.]{redd2011identification}
A.D. Redd, A.~Collinson-Streng, C.~Martens, S.~Ricklefs, C.E. Mullis,
  J.~Manucci, A.A.R. Tobian, E.J. Selig, O.~Laeyendecker, N.~Sewankambo, et~al.
\newblock Identification of {HIV} superinfection in seroconcordant couples in
  {Rakai}, {Uganda}, by use of next-generation deep sequencing.
\newblock \emph{J. clin. microbiol.}, 49\penalty0 (8):\penalty0 2859--2867,
  2011.

\bibitem[Steel(2005)]{steel2005phylogenetic}
M.~Steel.
\newblock Phylogenetic diversity and the greedy algorithm.
\newblock \emph{Syst. biol.}, 54\penalty0 (4):\penalty0 527--529, 2005.

\bibitem[Theodoridis and Koutroumbas(2006)]{patternRecognitionBook}
S.~Theodoridis and K.~Koutroumbas.
\newblock \emph{Pattern Recognition}.
\newblock Academic Press, San Diego, 3rd edition, 2006.

\bibitem[Villani(2003)]{villani2003topics}
C.~Villani.
\newblock \emph{Topics in optimal transportation}, volume~58.
\newblock American Mathematical Society, 2003.

\bibitem[Wu et~al.(2009)Wu, Hugenholtz, Mavromatis, Pukall, Dalin, Ivanova,
  Kunin, Goodwin, Wu, Tindall, et~al.]{wu2009phylogeny}
D.~Wu, P.~Hugenholtz, K.~Mavromatis, R.~Pukall, E.~Dalin, N.N. Ivanova,
  V.~Kunin, L.~Goodwin, M.~Wu, B.J. Tindall, et~al.
\newblock A phylogeny-driven genomic encyclopaedia of {Bacteria} and {Archaea}.
\newblock \emph{Nature}, 462\penalty0 (7276):\penalty0 1056--1060, 2009.

\bibitem[Yule(1925)]{yule1925mathematical}
G.U. Yule.
\newblock {A mathematical theory of evolution, based on the conclusions of Dr.
  JC Willis, FRS}.
\newblock \emph{Phil. Tran. Royal Soc. London (B)}, 213:\penalty0 21--87, 1925.

\bibitem[Zhang and Couloigner(2005)]{zhang2005new}
Q.~Zhang and I.~Couloigner.
\newblock A new and efficient k-medoid algorithm for spatial clustering.
\newblock \emph{Comp. Sci. and Its Appl.--ICCSA 2005}, pages 207--224, 2005.

\bibitem[Ziegler(1995)]{ziegler1995lectures}
G.M. Ziegler.
\newblock \emph{Lectures on polytopes}, volume 152 of \emph{Graduate Texts in
  Mathematics}.
\newblock Springer, 1995.

\end{thebibliography}
\bibliographystyle{plainnat}

\notforarxiv{\newpage}

\section{Appendix: A more complete description of ADCL minimization}

The distance $d(\cdot, \cdot)$ between points on a tree is defined to be the length of the shortest path between those points.
A \emph{rooted subtree} is a subtree that can be obtained from a rooted tree $T$ by removing an edge of $T$ and taking the component that does not contain the original root of $T$.
The \emph{proximal} direction in a rooted tree means towards the root, while the \emph{distal} direction means away from the root.
We emphasize that the phylogenetic trees here are considered as collections of points with distances between them, i.e.\ metric spaces, such that by a subset of a phylogenetic tree we mean a subset of those points.

\subsection{Introduction to ADCL}

\begin{defn}
  A \emph{mass map} on a tree $T$ is a Borel measure on $T$.
  A \emph{mass distribution} on a tree $T$ is a Borel probability measure on $T$.
\end{defn}

\begin{defn}
  Given a subset $X \subset L(T)$ of leaves and a mass distribution $\mu$, define the Average Distance to the Closest Leaf (ADCL) to be the expected distance of a random point distributed according to $\mu$ to its closest leaf in $X$. That is,
  \[
    \ADCL_\mu(X) = \EE\left[\min_{\ell \in X} \, d(P, \ell)\right].
  \]
  where $P \sim \mu$.
\end{defn}

\begin{prob}
  \label{prob:ADCL}
  Minimize ADCL for a given number of allowed leaves.
  That is, given $0 < \maxleaves \leq |L(T)|$ and probability measure $\mu$, find the $X \subset L(T)$ with $|X| = \maxleaves$ minimizing $\ADCL_\mu(X)$.
\end{prob}

The expected distance of a randomly sampled point $P \sim \mu$ to a fixed point $\ell$ is equal to the amount of work required to move mass distributed according $\mu$ to the point $\ell$, when work is defined as mass times distance.

\subsection{Voronoi regions}

In this section we connect the above description with the geometric concept of a Voronoi diagram.

\begin{defn}
  Given a subset $X \subset L(T)$ of leaves and $\ell \in X$, the \emph{Voronoi region} $V(\ell,X)$ for leaf $\ell$ is the set of points of $T$ such that the distance to $\ell$ is less than or equal to the distance to any other leaf in $X$.
  The \emph{Voronoi diagram} for a leaf set $X$ in the tree is the collection of Voronoi regions for the leaves in $X$.
\end{defn}

Note that the Voronoi regions by this definition are closed sets that are not disjoint; they intersect each other in discrete points where the distances to leaves are equal.

\begin{defn}
  Given a subset $Z$ of $T$, a mass distribution $\mu$ and a leaf $\ell$, let $\vwork_\mu(Z,\ell)$ be the work needed to move the mass of $\mu$ in $Z$ to the leaf $\ell$.
\end{defn}

The following simple lemma allows us to express the ADCL in terms of the Voronoi regions.

\begin{lem}
  \label{lem:ADCLvoronoi}
  Let $\mu$ be a mass distribution. Then
  \begin{equation}
    \ADCL_\mu(X)= \sum_{\ell \in X} \vwork_\mu(V(\ell,X),\ell).
  \end{equation}
\end{lem}

\begin{proof}

  For a given $p \in T$,
  \[
    \min_{\ell \in X} d(p,\ell) = \sum_{\ell \in X} 1_{V(\ell,X)}(p) d(p,\ell).
  \]
  where $1_{V(\ell,X)}$ is the indicator function for the set $V(\ell,X)$.
  Let $P \sim \mu$. Then
  \[
    \begin{split}
      \EE \left[ \min_{\ell \in X} d(P,\ell) \right] & = \int_{p \in T} \min_{\ell \in X} d(p,\ell) d \mu \\
        & = \sum_{\ell \in X} \int_{p \in V(\ell,X)} d(p,\ell) d \mu \\
        & = \sum_{\ell \in X} \vwork_\mu(V(\ell,X), \ell)
    \end{split}
  \]
  where the last step is by the optimization definition of the KR distance \citep{EvansMatsenPhyloKR12}.
\end{proof}

\subsection{Dynamic program}

\subsubsection{Background}

This section presents a full solution to Problem~\ref{prob:ADCL} via a dynamic program.
This dynamic program will descend through the tree selecting each rooted subtree $S$ in a depth first manner, and solving the optimization for every amount and direction of mass transport through the root of $S$.
Because the algorithm constructs every solution that is not sub-optimal, the algorithm is exact.

\subsubsection{Bubbles}

A first observation for the exact algorithm is that the tree can be divided into connected sets, that we call \emph{bubbles}, with the following property: irrespective of the set of leaves $X$ of the tree $T$ that are selected, any pair of points in the same bubble will share the same closest leaf in $X$.
Because of this characteristic, all that is needed to decide the fate of every particle of mass is to decide optimal mass transport on a per-bubble basis.
Indeed, if it is optimal to move mass at point $p$ to a leaf $\ell$, then the same must be true for every other point $q$ in in the bubble.
This observation turns the search for an optimal subset into an optimal assignment of bubbles to leaves of the tree, such that the total number of leaves that are assigned a bubble has cardinality at most $k$.
As described below, the partition of the tree into bubbles is in fact the refinement of all possible Voronoi diagrams for all subsets of the leaves (along with the partition by edges, which is put in for convenience).

Recall that partitions of a set are partially ordered by inclusion, such that $\calA \leq \calA'$ for two partitions $\calA$ and $\calA'$ iff every $V \in \calA$ is contained in a $V' \in \calA'$.
Partitions are a complete lattice with this partial order, thus there exists a greatest lower bound for any collection of partitions; define $\calA \wedge \calB$ be the greatest lower bound for any partitions $\calA$ and $\calB$.
In practice this means finding the ``coarsest'' partition such that pairwise intersections of sets in the partitions are represented: for example,
\[
  \{A, X \setminus A\} \wedge \{B, X \setminus B\} = \{A \cap B, A \setminus B, B \setminus A, X \setminus (A \cup B)\}.
\]
We will be interested in partitions of phylogenetic trees, and the boundaries of partitions can be thought of as ``cuts'' on edges or internal nodes.
Thus if $\calA$ and $\calB$ are two partitions of a tree $T$, $\calA \wedge \calB$ is the partition with every ``cut.''

Let $\calV(X)$ be the Voronoi diagram of $T$ for some subset $X$ of the leaves of $T$.
Let $\calE$ be the partition of $T$ such that the edges of $T$ are the sets of the partition.

\begin{defn}
  The \emph{bubble partition} of a tree $T$ is the coarsest partition refining all of the Voronoi decompositions of $T$ and the edge partition:
  \[
    \bubbles(T) := \calE \wedge \bigwedge_{X \subset L(T)} \calV(X)
  \]
\end{defn}

This partition forms the basis of our approach.
By Lemma~\ref{lem:ADCLvoronoi}, an exhaustive approach to Problem~\ref{prob:ADCL} would involve trying all Voronoi diagrams for a given tree and transporting the mass in each of the regions to the closest leaf.
However, $\bubbles(T)$ is the refinement of all Voronoi partitions.
Because every point in a bubble has the same closest leaf, every optimal solution can be completely described by deciding to what leaf the mass in $B$ gets sent for each $B \in \bubbles(T)$.
The number of bubbles is quadratic in the number of leaves irrespective of the given mass distribution.

In particular, by describing the optimization algorithm in terms of bubbles, it will work in the case of a continuous mass distribution.
What is needed is a way to calculate the amount of work needed to move a continuous mass distribution to one side of a bubble.
This can be done using a simple integral as described in \citep{EvansMatsenPhyloKR12}, however, the above-described \rppr\ implementation is in terms of a discrete distribution of mass.

\subsubsection{Recursion introduction}

In this section we will describe a recursion that will solve Problem~\ref{prob:ADCL} as described above.
Fix the number of allowed leaves $\maxleaves$.
The recursion is depth-first starting at a root, which can be arbitrarily assigned if the tree is not already rooted.
\emph{Partial solutions} start at the leaves, are modified and reproduce as they travel through bubbles, then get combined at internal nodes.
We remind the reader that a solution to Problem~\ref{prob:ADCL} is completely specified by the destination of the mass for each bubble.

These partial solutions will be denoted by labeled tuples, either $\RMD(\leafset, \cldist, \wksubtot)$ for a root mass distal solution, or $\RMP(\leafset, \cldist, \proxmass, \wksubtot)$ for a root mass proximal solution as follows.
The $\leafset$ component of the partial solution is the \emph{leaf set}: the leaves that have been selected for the partial solution to the ADCL problem.
The $\cldist$ component is the \emph{closest leaf distance}: the distance to the closest leaf in the partial solution.
The $\proxmass$ component is the \emph{proximal mass}: the mass that is moved to the root of $S$ by this partial solution (RMD solutions always have zero proximal mass).
The $\wksubtot$ component is the \emph{work subtotal}: the amount of work needed for the partial solution to move the mass in $S$ to either the root of $S$ or a leaf in $\leafset$.
Whether a partial solution is an RMP or an RMD solution defines its \emph{mass direction class}.

The depth first recursion will maintain a list of partial solutions that gets updated upon traversing bubbles and internal nodes.
We remind the reader that bubbles never span more than a single edge.

\subsubsection{Base case at a leaf}

There are two base cases at a leaf: that of not including the leaf and that of including the leaf.
The partial solution that corresponds to including the leaf $\ell$ is $\RMD(\{\ell\}, 0, 0)$ and that of not including the leaf is $\RMP(\emptyset, \infty, 0, 0)$.
From there, we move proximally through the bubbles along the edges and through the internal nodes as follows.
Note that these partial solutions at each leaf are then passed through the bubbles directly proximal to their leaf.

\subsubsection{Moving through a bubble}

Assume the algorithm is traversing a bubble along an edge, such that the edge length in the bubble is $\bublen$ and the amount of mass in the bubble is $\bubmass$.
Define $\wkdstl$ to be the amount of work required to move the mass in the bubble to the distal side of the bubble, and $\wkprox$ to be the corresponding amount of work for the proximal side.
We now describe the steps required to update this collection of partial solutions going from the distal to the proximal side of the bubble.

The first step is to update the existing partial solutions.
In this updating step the mass-direction class will not be changed (in the second step we will construct RMP solutions from RMD solutions).
RMD solutions get updated as follows:
\[
  \RMD(\leafset, \cldist, \wksubtot)
\]
maps to
\[
  \RMD(\leafset, \cldist + \bublen, \wksubtot + \wkdstl + \bubmass \cdot \cldist)
\]
as RMD solutions must move the mass of the current bubble to a leaf, and moving it to the closest selected leaf is optimal.

On the other hand, RMP solutions will have the mass of the current bubble moving away from the leaves of $S$.
Thus
\[
  \RMP(\leafset, \cldist, \proxmass, \wksubtot)
\]
maps to
\[
  \RMP(\leafset, \cldist + \bublen, \proxmass + \bubmass, \wksubtot + \wkprox + \proxmass \cdot \bublen).
\]

The second step is to consider solutions such that the mass transport on the distal and proximal sides of the bubble are not the same.
In that case, the optimal directions of mass movement on distal and proximal edges for a given bubble must be pointing away from each other; the alternative could not be optimal.
Thus here we consider adding an RMP solution based on a previous RMD solution.
This solution has all of the mass going to the leaves in $S$ except that of the current bubble, which moves proximally.
This step can be ignored if $\bubmass = 0$.
Given a $\RMD(\leafset, \cldist, \wksubtot)$ with $\leafset \neq \emptyset$, add the resulting $$\RMP(\leafset, \cldist + \bublen, \bubmass, \wksubtot + \wkprox)$$ to the list of possible solutions if $\wkprox < \wkdstl + \bubmass \cdot \cldist$.

The following simple lemma reduces the number of bubbles that must be considered.
\begin{lem}
  \pushQED{\qed}
  Given two neighboring bubbles on a single edge, such that the proximal bubble has zero mass.
  The recursive step in this section after progressing through these two bubbles is identical to one where the two bubbles are merged.
  \popQED
\end{lem}

\subsubsection{Moving proximally through an internal node}

When encountering an internal node, the algorithm first combines all tuples of partial solutions for each subtree as follows.
Assume we are given one $\PS_i$ from each of the subtrees, where $\leafset_i$, $\cldist_i$, $\proxmass_i$, and $\wksubtot_i$ are as above for the $i$th tree (define $\proxmass_i = 0$ for RMD solutions).

At least one, and possibly two, partial solutions can be constructed from the partial solutions in the subtrees.
There is always one solution where the proximal mass, if it exists, continues moving away from the leaves of $S$; we will call this the ``continuing'' solution.
Sometimes it is also possible for the proximal mass to go to a leaf in one of the subtrees, giving another solution we will call the ``absorbing'' solution.

The continuing solution in the case that all solutions are RMD solutions is
\[
  \RMD \left(\bigcup_{i} \leafset_i, \min_i \cldist_i, \sum_i \wksubtot_i \right).
\]
If any of the solutions are RMP solutions, then the resulting solution is
\[
  \RMP \left(\bigcup_{i} \leafset_i, \min_i \cldist_i, \sum_i \proxmass_i, \sum_i \wksubtot_i \right)
\]
where $\pi_i=0$ for RMD solutions.

Now, if at least one leaf is selected in one of the subtrees, then it could be optimal to move the proximal mass to the closest leaf of the existing RMD solutions to make an absorbing solution; in that case we also have the new RMD solution
\[
  \RMD \left(\bigcup_{i} \leafset_i, \min_i \cldist_i, \sum_i \wksubtot_i + \left(\min_i \cldist_i \right) \cdot \sum_i \proxmass_i \right).
\]

Given an internal node with $k$ subtrees, combine all partial solutions of the subtrees in this way.
The complete collection of solutions given a set of $\PSset_i$ for each subtree is the union of this process applied to every element of the Cartesian product of the $\PSset_i$.
Throw away any partial solutions such that the cardinality of the $\leafset$ is greater than $\maxleaves$.

\subsubsection{Termination at the root}

Select the RMP solution $\RMP(\leafset, \cldist, \wksubtot)$ with $|\leafset| = \maxleaves$ with the smallest $\wksubtot$.
Note that in fact all solutions for number of chosen leaves less than $k$ are generated by this algorithm.

\subsubsection{Avoiding computation of suboptimal solutions}

By na\"ively combining all of the solutions described above we may get solutions that cannot be optimal for any structure of the rest of the tree.
These solutions greatly reduce the speed of the algorithm when carried along as described above.
In this section we describe a way to avoid these solutions using geometry (Fig.~\refFIGlowerHull).

The culling strategy employed here is to eliminate partial solutions that would not be optimal for any amount and direction of mass transport through the root of $S$.
This is achieved by first binning the partial solutions by the number of leaves $k$ they employ then further binning them by mass-direction class.

For an RMP solution the amount of work required in a given partial solution to move the mass in $S$ to some selected leaf is equal to the mass transport within $S$ plus the amount of work required to move the proximal mass of that partial solution to some leaf proximal to $S$.
Imagine that this proximal leaf is distance $x$ away from the root of $S$.
For a given partial solution we can plot the amount of work to move the mass in $S$ to some selected leaf with respect to $x$.
This will be $\wksubtot + x \cdot \proxmass$.

Similar logic applies for RMD solutions but where the appropriate parameter $x$ to use is the amount of mass that comes proximally through $\troot$.
The total amount of work in that case then, is $\wksubtot + x \cdot \cldist$.

These considerations motivate the following definition.

\begin{defn}
  The \emph{subwork} $f_\PS$ for a partial solution $\PS$ is the function
  \[
    f_\PS(x) =
    \begin{cases}
      \wksubtot + x \cdot \proxmass & \hbox{if}\ \PS = \RMP(\leafset, \cldist, \proxmass, \wksubtot) \\
      \wksubtot + x \cdot \cldist & \hbox{if}\ \PS = \RMD(\leafset, \cldist, \wksubtot)
    \end{cases}
  \]
  Let the $x$ in $f_\PS(x)$ be called the \emph{subwork parameter}.
\end{defn}

\begin{defn}
  Assume $\PSset$ is a set of partial solutions with the same number of leaves and the same root mass direction.
  The \emph{optimality interval} $I_\PSset(\PS)$ is the interval for which $\PS$ is optimal compared to the other solutions in $\PSset$, namely
  \[
    I_\PSset(\PS) = \left\{ x \in [0,\infty) : f_\PS(x) \leq f_{\PS'}(x) \forall \PS' \in \PSset \right\}
  \]
\end{defn}
It can be easily seen that the set defined in this way is actually an interval.
We can ignore partial solutions that have an empty optimality interval.

These optimality intervals can be found by using the double description algorithm as described in the algorithm introduction.
Specifically, a line with equation $y=mx+b$ will get translated into the half-plane constraint $y \leq mx+b$.
The set of points in the plane that satisfy this collection of inequalities is called a \emph{convex polytope}.
A convex polytope can be equivalently described as the intersection of a number of half-spaces (a so-called \emph{H-description}) or as the convex hull of a number of points (a so-called \emph{V-description}).
In this setting the set of $x$ values between vertices are the optimality intervals, and the lines that contact pairs of neighboring vertices correspond to partial solutions that are optimal for some value of $x$ (Fig.~\refFIGlowerHull).

The ``double description'' algorithm \citep{motzkin1983double,fukuda1996double} is an efficient way to go from an H-description to a V-description which also collects information on what linear constraints contact what facets.
One way to use this perspective on optimal solutions is to simply throw away partial solutions that have empty optimality intervals after combining.
Another is to use optimality intervals to guide the combination of solutions in at internal nodes as described in the next section.

We perform a pre-filtering step before running the double description algorithm to discard lines that could never be facets.
Denote lines as being pairs of $(m, b)$, where $m$ is a slope and $b$ is a $y$-intercept.
Clearly if $m_1 < m_2$ and $b_1 < b_2$ then the line $(m_1,b_1)$ will lie below $(m_2,b_2)$ for all positive $x$.
Therefore $(m_2,b_2)$ will never be a facet and can be ignored.
We quickly eliminate some of these clearly suboptimal partial solutions by sorting the $(m,b)$ pairs in terms of increasing $m$.
If in this ordering, the line $(m_1,b_1)$ precedes $(m_2,b_2)$ with $b_2 \geq b_1$ then we discard $(m_2,b_2)$.

We can also include some global information as inequalities.
For RMP solutions the subwork parameter is bounded between the closest and farthest leaves in the proximal part of the tree.
For RMD solutions the subwork parameter is bounded between zero and the total amount of mass in the proximal part of the tree.
These additional constraints further cut down optimality intervals and reduce the number of solutions.

\subsubsection{Optimality intervals and solution combination}
\label{sec:combination}

Here we explore ways of using optimality intervals to reduce the number of partial solutions that must be combined between subtrees.
We will describe the partial solutions combination in terms of combining pairs of subtrees.
If a given internal node has more than two subtrees, say $T_1, \dots, T_k$, then we can combine over partial solutions for $T_1$ and $T_2$, then combine those results with $T_3$, and so on.

Assume we are given two partial solutions $\PS_1$ and $\PS_2$, and these have two optimality intervals $I_1 = (l_1,u_1)$ and $I_2 = (l_2,u_2)$, respectively.
The criteria used to check if a given combination could be optimal depend on the mass-direction class, and we will describe the criteria on a case by case basis.
Let $\PS$ be the solution that is formed from the combination of $\PS_1$ and $\PS_2$, and denote its optimality interval with $I$.

If $\PS_1$ and $\PS_2$ are both RMP solutions, the subwork parameter for each will be the distance to the closest leaf in the proximal part of the tree.
Since the combination will also be an RMP solution, to be optimal for a subwork parameter $x$ each partial solution will have to be optimal for that $x$.
Thus $I = I_1 \cap I_2$, and $\PS$ is viable if $I \neq \emptyset$.

If $\PS_1$ and $\PS_2$ are both RMD solutions, the subwork parameter is the amount of mass that comes proximally through the root of $S$.
When mass comes proximally through the root, it will go to the subtree that has the closest leaf.
For this reason, $I$ will be the optimality interval of the partial solution with the smallest $\cldist$.
If the subtrees have identical $\cldist$'s, then $I$ is the smallest interval containing $I_1$ and $I_2$.

Recall that when one partial solution is an RMP solution and the other is an RMD solution, then we can get either an RMP or an RMD solution.
Assume first $\PS_1$ is an RMP solution, $\PS_2$ is an RMD solution, and $\PS$ is an RMP solution.
Since $\PS$ is an RMP solution, no mass will be sent into $T_2$ from outside of it.
Thus $\PS$ should only be used if $\PS_2$ has the smallest $\wksubtot$ across all partial solutions that have the same number of leaves as $\PS_2$.
Also, the solution could be optimal only if $\cldist_2$ is greater than the upper bound of $I_1$, because otherwise it would be optimal to send the proximal mass of $T_1$ proximally into $T_2$.
If $\PS_1$ is an RMP solution, $\PS_2$ is an RMD solution, and $\PS$ is an RMD solution, the subwork parameter for $\PS$ is the amount of mass coming from above.
Because the proximal mass of $\PS_1$ will be sent into $T_2$, the optimality interval $I$ is $I_2 \cap [\proxmass_1,\infty)$.
The corresponding partial solution is valid if $I$ is nonempty and $\cldist_2 \in I_1$.

\notforarxiv{
\newpage
\section{Legends}
\FIGmassTransport
\FIGlocalMinimum
\TABlocalMinimum
\FIGrmdRmp
\FIGlowerHull
\FIGpamOptimality
\FIGkTimeComplexity
\FIGnTimeComplexity
\FIGchimera
}

\end{document}